\documentclass[
  prb,
  twoside,
  twocolumn,
  showpacs,
  floatfix,
  superscriptaddress,
  10pt,
  preprintnumbers,
  citeautoscript,
]{revtex4-2}

\usepackage{amsmath}
\usepackage{comment}
\usepackage{graphicx}
\usepackage[utf8]{inputenc}
\usepackage{times}
\usepackage{mathtools}
\usepackage{physics}
\usepackage{colortbl}
\usepackage{array}
\usepackage{dcolumn}
\DeclareUnicodeCharacter{2212}{-}
\usepackage{amssymb}
\usepackage{amsmath}
\usepackage{units}
\usepackage{xcolor,colortbl}
\usepackage[colorlinks=true]{hyperref}

\hypersetup{
    unicode=false,          
    pdftoolbar=true,        
    pdfmenubar=true,        
    pdffitwindow=false,     
    pdfstartview={FitH},    
    pdfnewwindow=true,      
    colorlinks=true,        
    linkcolor=blue,         
    citecolor=blue,         
    filecolor=blue,         
    urlcolor=blue           
}

\usepackage{xcolor}
\usepackage{booktabs}
\AtBeginDocument{
\heavyrulewidth=.08em
\lightrulewidth=.05em
\cmidrulewidth=.03em
\belowrulesep=.65ex
\belowbottomsep=0pt
\aboverulesep=.4ex
\abovetopsep=0pt
\cmidrulesep=\doublerulesep
\cmidrulekern=.5em
\defaultaddspace=.5em
}

\newcommand{\chg}[1]{{\color{black} #1}}

\newcommand{\mono}[1]{$\mathrm{C}^{#1}_{\mathrm{1i}}$}
\newcommand{\di}[1]{$\mathrm{C}^{#1}_{\mathrm{2i}}$}
\newcommand{\tri}[1]{$\mathrm{C}^{#1}_{\mathrm{3i}}$}
\newcommand{\tetra}[1]{$\mathrm{C}^{#1}_{\mathrm{4i}}$}
\newcommand{\penta}[1]{$\mathrm{C}^{#1}_{\mathrm{5i}}$}
\newcommand{\hexa}[1]{$\mathrm{C}^{#1}_{\mathrm{6i}}$}

\begin{document}
\title{From Mono- to Hexa-Interstitials: Computational Insights into Carbon Defects in Diamond}

\newcommand{\Wigner}{HUN-REN Wigner Research Centre for Physics, P.O. Box 49, H-1525 Budapest, Hungary}
\newcommand{\BME}{Department of Atomic Physics, Institute of Physics, Budapest University of Technology and Economics, M\H{u}egyetem rakpart 3., H-1111 Budapest, Hungary}
\newcommand{\MTA}{ MTA-WFK Lendület “Momentum” Semiconductor Nanostructures Research Group, P.O. Box 49, H-1525 Budapest, Hungary}

\author{Nima Ghafari Cherati}
\affiliation{\Wigner}
\affiliation{\BME}

\author{Arsalan Hashemi}
\affiliation{European Laboratory for Learning and Intelligent Systems (ELLIS)  Institute Finland, Maarintie 8, 02150 Espoo, Finland}

\affiliation{Department of Technical Physics, University of Eastern Finland, P.O. Box 1627, FI-70211 Kuopio, Finland}

\author{\'Ad\'am Gali}
\affiliation{\Wigner}
\affiliation{\BME}
\affiliation{\MTA}
\email{gali.adam@wigner.hun-ren.hu}

\begin{abstract}
We present a comprehensive first-principles investigation of carbon self-interstitial defects in diamond, ranging from mono- to hexa-interstitial complexes.
By quantum mechanical density functional theory, empowered by interatomic potential models, we efficiently sample the complex configurational landscape and identify both known and previously unreported defect geometries.
Our results reveal a pronounced energetic driving force for aggregation: the formation energy per interstitial decreases systematically from isolated split interstitials to compact multi-interstitial clusters, with the tetra-interstitial platelet emerging as a particularly stable structural motif.
Additionally, charge analysis indicates that the predominantly covalent bonding in diamond becomes more polar within the defect centers.
Analysis of defect energy levels shows that only the investigated mono-, di-, penta-, and hexa-interstitial complexes introduce in-gap electronic states, whereas the tri- and tetra-interstitial clusters are electronically inert.
Vibrational spectroscopies further reveal that self-interstitials generate characteristic signatures.
Short carbon–carbon bonds inside the defect cores give rise to high-frequency vibrational modes between 1375 and 1925 cm$^{-1}$, which are strongly IR-active but exhibit weak Raman activity.
\chg{Through a systematic analysis of metastable configurations, we identify the 3H defect center as a neutral di-interstitial defect. Based on this identification, we further suggest that the TR12 center may arise from a 3H-containing defect like a metastable hexa-interstitial configuration.}
Taken together, these findings provide a coherent picture of the structural, electronic, and vibrational characteristics of carbon self-interstitials and establish a robust framework for their experimental identification.
\end{abstract}

\maketitle

\section{Introduction}
One of the long-standing challenges in materials engineering is the controlled growth of truly pristine diamond.
Even the highest-purity single-crystal diamonds produced by modern high-pressure–high-temperature or chemical vapor deposition methods contain both extrinsic and intrinsic defects, including nitrogen (N), hydrogen (H), boron (B), carbon (C) vacancies, and self-interstitials, typically at ppm levels~\cite{Kalish2007,Nemanich2014}.
Remarkably, the standard classification scheme is itself defined by defects: type IaA, IaB, and Ib diamonds incorporate different configurations of N impurities; type IIa is essentially free of both N and B impurities, whereas type IIb contains B as the dominant impurity~\cite{Zhang2024,Misra2023,Zaitsev2020,breeding2009}.

However, defects do not always deteriorate the properties of diamond; on the contrary, they can endow it with technologically valuable functionalities~\cite{Rovny2022,Wolfowicz2021,Zhang2020}.
Among the hundreds of reported centers, the prototypical N-vacancy (NV) defect stands out for its optically detected magnetic resonance (ODMR) and millisecond-scale spin coherence at room temperature, which together underpin its powerful quantum sensing capabilities~\cite{Balasubramanian2008}.
In contrast to the NV center, which has been extensively studied over the past decade and measures only the projection of the magnetic field along its single symmetry axis, diamond also hosts a multi-orientation center, TR12, that couples to the field along twelve symmetry-related axes.
This enables true vector magnetometry with wide angular coverage from a single ensemble spectrum, allowing for the reconstruction of the full magnetic field without sample rotation and enhancing robustness to arbitrary field orientations~\cite{Foglszinger2022}.
Despite its significance, the atomistic structures of TR12 and several putative C self-interstitial complexes, including the 3H and R-family centers~\cite{Iakoubovskii2000,Loubser1978}, remain unresolved.

In the ongoing effort to identify the atomistic structure of TR12, numerous experimental investigations have converged on the view that TR12 is an intrinsic interstitial-related defect in type IIa diamond.
Following the first report of TR12 in 1956~\cite{Clark1956}, Walker \textit{et al.} showed that the center can be generated in high-purity type IIa diamond by electron irradiation, without introducing foreign impurities, implying that only lattice atoms participate in its formation~\cite{Walker1977}.
Its nonlinear dose dependence further points to a multi-particle intrinsic complex rather than a simple point defect.
Optical studies subsequently established TR12 as a distinct vibronic system with a zero-phonon line (ZPL) near 2.63 eV, observed in nitrogen-free diamond after damage, consistent with an aggregated intrinsic damage center rather than an impurity-related defect~\cite{Davies1981}.
Annealing experiments then provided a mechanistic insight: while several self-interstitial-related centers anneal out in the temperature window below vacancy mobility ($\sim 900$ K), the TR12 (and 5RL) bands grow in the same range, indicating that TR12 is fed by mobile carbon self-interstitials released from precursor complexes~\cite{Allers1998}.

Ion-implantation studies strengthened this picture and highlighted TR12's photophysical promise, reporting optically active centers with a Debye–Waller factor of order 0.1 at cryogenic temperatures, higher than that of the NV center, suggesting improved photon emission into the ZPL~\cite{Naydenov2009}. Spatially resolved hyperspectral mapping further showed that the TR12 distribution along an implantation track can be treated as a \emph{split self-interstitial map}, with TR12 produced throughout much of the profile before and after high-temperature annealing and suppressed in the most heavily damaged end-of-range region, supporting an association with split-interstitial-derived configurations~\cite{Turaga2019}.
Single-center ODMR measurements then established TR12 as a spin-active defect with multi-orientation response and a level structure involving singlet states coupled via a long-lived metastable triplet, enabling vector magnetometry without sample rotation and with enhanced robustness to arbitrary field directions~\cite{Foglszinger2022}.
Most recently, focused carbon-ion microbeam experiments directly correlated depth-resolved TR12 formation with the local interstitial reservoir and competition with the 3H center, supporting TR12 as a self-interstitial cluster defect that can be engineered for high-performance magnetometry~\cite{Chai2025}.

Despite the wealth of compositional evidence, pinpointing the exact atomistic structures of defect centers in bulk diamond remains exceptionally challenging, because nearly all available experimental probes provide only indirect signatures rather than direct real-space structural information.
Techniques such as optical spectroscopy, electron paramagnetic resonance (EPR), ODMR, and infrared or Raman scattering constrain only the symmetry, spin state, charge state, and local vibrational characteristics.
As a result, many distinct microscopic configurations can produce nearly identical experimental fingerprints in these observables~\cite{Davies1994,Newton2002}.
Moreover, bulk diamond is both extremely stiff and dense, rendering direct real-space imaging of buried point defects at atomic resolution effectively impossible.
Compounding the difficulty, many relevant centers occur at low concentrations and can exhibit multiple charge states or dynamic Jahn-Teller distortions, which further broaden or obscure their spectral signatures~\cite{Davies1994,Foglszinger2022}.

In this context, density functional theory (DFT) provides a powerful complementary route: candidate defect geometries can be constructed, relaxed at the atomistic level, and evaluated through their formation energies, charge-transition levels, and local relaxations. Importantly, DFT can also predict directly comparable observables, including ZPL energies, zero-field splitting (ZFS) parameters, and local vibrational modes, enabling stringent, quantitative comparisons with experiment~\cite{Breuer1995,Goss2003,Gali2019}.
Comparison between computed and measured signatures can exclude broad classes of structural models and, in favorable cases, support a unique microscopic assignment.
For more complex centers, such as TR12, ongoing DFT efforts are steadily narrowing the landscape of viable carbon self-interstitial complexes consistent with their spectroscopic fingerprints~\cite{Goss2003,Foglszinger2022}.

However, early DFT studies, predominantly based on traditional semilocal functionals~\cite{PerdewPBE1996,Perdew1981LDA}, showed that the tetrahedral, hexagonal, and bond-centred mono-interstitial configurations are highly unstable, whereas the $\langle 001\rangle$ split interstitial is the only robust elementary self-interstitial, with a large formation energy and a central role in self-diffusion~\cite{Bernholc1988PRL,Breuer1995}.
In a complementary line of work, Goss and co-workers investigated di-, tri-, and tetra-interstitial clusters, proposing these small aggregates as building blocks for larger planar interstitial structures (platelets) and assigning several configurations to specific triplet EPR centres, including R2, R1, and O3~\cite{Goss2000JPCM, Goss2001DRM, Goss2001PRB, Goss2003}.
More recent hybrid-functional studies have refined the vibrational and EPR signatures of $\langle 100\rangle$ interstitials, their N-containing analogues and vicinal double-interstitial complexes, and have also examined selected vacancy–interstitial pairs and their recombination pathways~\cite{Colasuonno2020, Komarovskikh2024, Salustro2016JCP}.
Yet, these calculations have been limited to the previously reported high-symmetry configurations, typically with no more than four interstitial atoms.
This gap motivates the present work: a comprehensive, high-accuracy screening of interstitial complexes as microscopic models for TR12.

In this work, we perform a systematic first-principles investigation of carbon self-interstitial defects.
Motivated by suggestions that TR12 may involve aggregates of up to six carbon self-interstitials~\cite{Mainwood1994}, we extend our analysis to clusters containing one to six self-interstitial atoms \chg{that can still be treated accurately with first-principles methods in a computationally feasible model system}.
We first investigate the structural properties of these configurations and then compare their relative formation energies to identify the most stable defects.
We also investigate the electronic and vibrational properties of the most stable complexes, providing Raman and IR spectroscopic fingerprints that enable the identification of electronically and optically silent configurations. \chg{Furthermore, we assess candidate carbon-interstitial-related color centers in the visible range, including the TR12 center, by comparing computed photoluminescence spectra with experiment.}

\section{Methodology}
\label{sec:methodology}
DFT calculations were performed using VASP~\cite{kres1} within the projector-augmented-wave (PAW) framework~\cite{PAW1994}.
The wave functions were expanded in a plane-wave basis with a kinetic-energy cutoff of 370~eV.
Total energies and electronic-structure properties were evaluated using the screened hybrid Heyd-Scuseria-Ernzerhof (HSE06) functional~\cite{Paier2006HSE}. For HSE06, all structures were fully relaxed until the total energy difference between successive ionic steps was below $10^{-5}$~eV and the forces on each atom were less than $10^{-2}$~eV/\AA.
For vibration assessment, the computationally cheaper Perdew-Burke-Ernzerhof (PBE) functional~\cite{PerdewPBE1996} was used.
To this end, the total energies in both geometry relaxation and in phonon calculations were converged within $10^{-6}$~eV and the forces to within  $10^{-3}$~eV/\AA.
We verified that no imaginary phonon modes were present, confirming that all structures were fully optimized and dynamically stable.
The optimized lattice constant is $a=3.57$~\AA\ (PBE) and $3.54$~\AA\ (HSE06), compared with the experimental $3.57$, and HSE06 reproduces a 5.4~eV band gap within 0.1~eV of experiment~\cite{ashcroft1976solid}.

Defect model systems were constructed in a $4\times4\times4$ cubic supercell containing 512 carbon atoms, derived from the conventional diamond unit cell.
With this supercell size, Brillouin-zone sampling at the $\Gamma$ point provides converged defect energetics and electronic properties.
During structural relaxations, the lattice vectors were kept fixed to isolate local distortions induced by the defects and to avoid spurious cell-wide strain effects.
Within this framework, the carbon number density increases from $3.522$ to $3.556$~g/cm$^{-3}$ when going from mono- to hexa-interstitial carbon complexes.

To easily generate more realistic initial configurations for carbon self-interstitial clusters in diamond, we employed a geometric Frenkel-pair construction carried out directly in configuration space. A spherical cavity of radius \(R\) was created at the center of a 512 atom diamond supercell by removing all atoms within a distance \(R\) from the cell center. The cavity radii (1.6--3.9~\AA) were selected based on the first several coordination shells of diamond, enabling a systematic modeling of voids capable of accommodating mono- to multi-interstitial complexes.
Within this cavity, \(m\) additional carbon atoms (\(1 \le m \le 6\)) were inserted using rejection Monte-Carlo sampling with a hard-sphere exclusion distance of 1.2~\AA, ensuring physically reasonable C$-$C separations.
Candidate positions were drawn from an isotropic, volume-uniform distribution, allowing the initial aggregates to sample a broad range of compact and extended motifs.
For each cluster size and $R$, 200 statistically independent initial structures were generated.
\chg{By deliberately sampling high-energy, distorted, and non-equilibrium configurations, we ensure adequate exploration of the defect configurational space.}

These configurations provide realistic starting points for subsequent geometry optimizations using a pre-trained neuroevolution potential (NEP) model~\cite{Fan2021} and an empirical interatomic potential~\cite{Erhart2005}, as implemented in GPUMD~\cite{Fan2022}.
The NEP model developed by Wang \textit{et al.}~\cite{wang_prb_2025} has been validated for carbon systems spanning mixed $sp^{2}$/$sp^{3}$ C$-$C bonding carbon-based materials and a broad range of densities, from slightly perturbed diamond-like structures to high-energy amorphous phases.
Although the NEP model performs well in many cases, particularly for metastable configurations, some geometries are better captured by an empirical potential, which we use to complement our dataset.
These pre-optimizations are performed in GPUMD under the $NVT$ ensemble with a Berendsen thermostat~\cite{Berendsen1984} at 1~K for $5 \times 10^{2}$ steps using a 0.5~fs time step.
Each structure is first rapidly pre-optimized using the interatomic potential, enabling efficient identification of low energy geometries and substantially reducing the number of costly, high precision DFT calculations required for final screening.

The stability of various defect forms was evaluated based on the calculated formation energies. Specifically, we computed the formation energy $E^q_\text{f}$ for  defects in a charge state $q$ as follows~\cite{Northrup1996}:
\begin{equation}
\label{eq:Eform}
    E^q_\text{f} = E^q_\text{tot} - E_{bulk} - \sum_{i} n_c \mu_c + qE_\text{Fermi} + E^q_\text{corr},
\end{equation}
where $ E^q_\text{tot}$ and $ E_\text{bulk}$  are the total energies of defective and pristine diamond, respectively, $ \mu_c$ is the chemical potential of carbon. $n_c$ is a number of added carbons as interstitials to the bulk system.   
$E_\text{Fermi}$ is the Fermi energy, referenced to the valence band maximum (VBM). $ E^q_\text{corr}$ is a correction term for the total energies of charged supercells~\cite{Freysoldt2018}.
The chemical potential of carbon was calculated from the diamond reference as the total energy per carbon atom.

The LOBSTER package~\cite{Nelson2020} was used to assess defect bonding strength via the integrated crystal orbital Hamilton population (ICOHP).
Within this framework, more negative ICOHP values correspond to stronger, more bonding interactions between the two carbon atoms, whereas less negative (or positive) values indicate weaker or antibonding interactions.

The Raman activity of a phonon mode $m$ with frequency $\omega_{m}$ and eigenvector $\mathbf{v}_{m}$ is determined by the derivative of the polarizability tensor $\chi$ with respect to the $\mathbf{v}_{m}$.
Within DFT calculations, this quantity can be equivalently evaluated from the derivative of the macroscopic dielectric tensor $\varepsilon_{\rm mac}$ as
\begin{equation}
R_{m} \propto \frac{\partial \chi}{\partial \mathbf{v}_{m}}
= \frac{\partial \varepsilon_{\rm mac}}{\partial \mathbf{v}_{m}} .
\label{eq:raman_activity}
\end{equation}
The non-resonant first-order Raman intensity is expressed as
\begin{equation}
I^{R}_m \propto \bigl| \mathbf{e}_s . \cdot R_m \cdot \mathbf{e}_i \bigr|^{2},
\label{eq:ir_activity}
\end{equation}
where $\mathbf{e}_{i}$ and $\mathbf{e}_{s}$ denote the polarization vectors of the incident and scattered light, respectively.
Spatial averaging over all possible scattering geometries yields the unpolarized Raman intensity used in this work.

The first-order infrared (IR) intensity is obtained from the derivative of the total dipole moment $\boldsymbol{\mu}$ with respect to the mode eigenvector,
\begin{equation}
I^{\rm IR}_m \propto
\biggl| \frac{\partial \boldsymbol{\mu}}{\partial \mathbf{v}_m} \biggr|^{2}.
\end{equation}
\chg{
To evaluate the Raman and IR activities according to each vibrational mode, we compute
Eqs.~\ref{eq:raman_activity} and \ref{eq:ir_activity} by displacing the atoms
along both the positive and negative directions defined by the phonon eigenvectors. This procedure requires $6N-6$ single-point calculations, where $N$ is the number of atoms in the supercell.

Phonon modes were obtained using the finite-difference method as implemented
in VASP. The resulting eigenvectors were mass-scaled to yield normalized
displacement vectors. The displacement amplitude was set to $0.015$~\AA.}
To generate continuous spectra, all discrete mode intensities were convoluted using Gaussian functions with a broadening of 5~cm$^{-1}$.

\chg{
Photoluminescence (PL) line shape of the defects was evaluated within the Huang--Rhys (HR) formalism~\cite{Huang1950, Alkauskas2014}.
This approach requires (i) the adiabatic potential energy surfaces of the ground and excited electronic states and (ii) the vibrational normal modes that couple to the optical transition. The excited-state geometry was obtained using the constrained-occupation $\Delta$SCF method as was explained for nitrogen-vacancy center in diamond~\cite{Gali2009}.
In this method, the electronic configuration of the target excited state is enforced by constraining the occupations of the Kohn-Sham (KS) orbitals, and the ionic positions are then relaxed in exactly the same way as for the ground state. This yields the minima of the potential-energy surfaces in both the ground state and the excited state. Denoting the corresponding total energies by \(E_\mathrm{g}^{\min}\) and \(E_\mathrm{e}^{\min}\), the zero-phonon-line (ZPL) energy is obtained as $E_\mathrm{ZPL} = E_\mathrm{e}^{\min} - E_\mathrm{g}^{\min}$.
The relaxed geometries were then used as reference structures for the analysis of phonon modes that participate in the optical transition.
The electron-phonon interactions can be formalized using the partial Huang-Rhys factor (HRF) as described below
\begin{equation}
    S_\lambda = \omega_\lambda Q_\lambda^2 \bigg/ 2\hbar,
\end{equation}
where $\hbar$ is the reduced Planck constant and the configurational coordinate $Q_\lambda$ for an optical process is defined as
\begin{equation} \label{eq:deltaq}
 Q_{\lambda} = \sum_{\alpha} \sqrt{m_{\alpha}} \langle(\bf{R}_{e, \alpha} - {\bf{R}}_{g, \alpha}) | {\bf{u}}_{\alpha, \lambda}\rangle. 
\end{equation}
Here, ${\bf{R}}_{g}$ and ${\bf{R}}_{e}$ are ground and excited state atomic coordinates,
while ${\bf{u}}_{\alpha, \lambda}$ indicates the normalized displacement vector corresponding to mode $\lambda$ with frequency $\omega_{\lambda}$ and $m_{\alpha}$ is mass of atom $\alpha$.
The total HRF and $Q^{2}$ are defined as $\rm{HRF} = \sum S_\lambda$ and $ Q^{2} = \sum Q^{2}_\lambda$, respectively.
They provide measures for the average number of phonons involved in the emission and the difference between the initial and final state geometries.
From the HRF, the DWF can be defined as $\text{DWF}=\exp(-\rm{HRF})$.

The spectral function that underlies the computation of the PSB is
\begin{equation}\label{eq:s}
S(\omega) = \sum_{\lambda}S_\lambda\delta(\omega - \omega_{\lambda}), 
\end{equation}
where the $\delta$-function is replaced by Gaussian with a broadening of \unit[3]{meV}.
Once the electron-phonon spectral function is computed,
the spectral distribution function can be determined as
\begin{equation}
    A(E_{\mathrm{ZPL}}/\hbar - \omega) = \int dt \exp(S(0)-S(t))\exp(i\omega t - \gamma|t|),
\end{equation}
where $S(t)$ is the Fourier transform of $S(\omega)$ and $\gamma$ determines the lifetime broadening of individual
contributions.
The PL intensity is proportional to $\omega^3 A(\omega)$.

}

\section{Results and discussion}
\label{sec:results}
In the remainder of this paper, defect complexes are labeled according to the number of interstitial carbon atoms they contain. mono- to hexa-interstitial clusters are denoted as $\mathrm{C}^{n}_{\rm{1i}}$, $\mathrm{C}^{n}_{\rm{2i}}$, $\mathrm{C}^{n}_{\rm{3i}}$, $\mathrm{C}^{n}_{\rm{4i}}$, $\mathrm{C}^{n}_{\rm{5i}}$, and $\mathrm{C}^{n}_{\rm{6i}}$, where the subscript $n$ labels distinct isomers within each cluster size.

\subsection{Structural Properties}
\label{subsect:str_props}

\begin{figure*}[ht!]
\centering
    \includegraphics[scale=1.65]{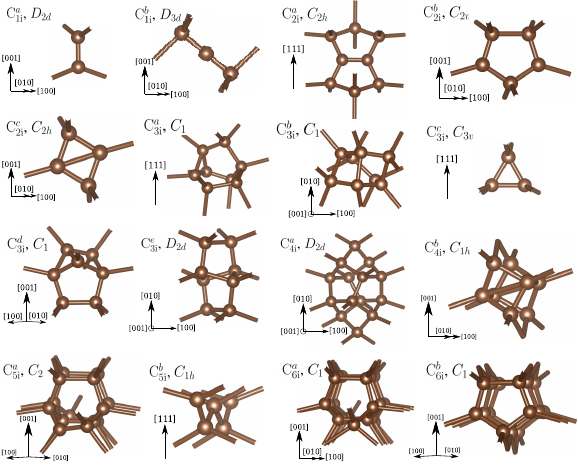}
    \caption{\label{fig:geo_inter} Optimized geometries of carbon self-interstitial configurations, ordered by increasing energy, together with their corresponding structural symmetries. Panels: \mono{a} and \mono{b} mono-carbon interstitials; \di{a} - \di{c} di-carbon; \tri{a} - \tri{e} tri-carbon; \tetra{a} and \tetra{b} tetra-carbon; \penta{a} and \penta{b} penta-carbon; and \hexa{a} and \hexa{b} hexa-carbon.}
\end{figure*}
During annealing, defects with low formation energies are expected to be stable, although metastable configurations can also persist when the energy barriers to their annihilation are sufficiently high.
Therefore, we applied an energy threshold, retaining as metastable candidates only those structures whose DFT-calculated relative formation energy lies within 4 eV of the lowest-energy configuration.
Structures exceeding this threshold are unlikely to form or survive under the annealing conditions considered. The resulting optimized geometries are shown in Fig.~\ref{fig:geo_inter}, ordered by increasing energy.

For the \mono{} defects, two distinct configurations were identified:
$\langle 001\rangle$-oriented split-interstitial ($I_1^{\langle001\rangle}$) and bond-centered defect (cf. Fig.\ref{fig:geo_inter} \mono{a} and \mono{b}, respectively).

Our HSE06 calculations indicate that the neutral \(I_1^{\langle001\rangle}\) in diamond has a singlet (\(S=0\)) ground state, with a triplet (\(S=1\)) state lying approximately $20$ meV higher in energy.
\chg{In fact, this energy difference equals 40 meV when the von Barth procedure is employed~\cite{VonBarth1972}, which avoids spin-state mixing.}
These findings agree with the previous theoretical studies~\cite{Goss2001PRB,Goss2001DRM,Goss2000JPCM}.
In addition, experimental measurements have identified a defect, known as the R2 center, exhibiting EPR activity, which has recently been shown through ZFS calculations to correspond to the $I_1^{\langle001\rangle}$ configuration~\cite{Komarovskikh2024}.
Its temperature-dependent EPR intensities place the triplet roughly $50$ meV above the singlet ground state, in close agreement with our result~\cite{Hunt2000PRB}.

We find that the structural symmetry of $I_1^{\langle 001 \rangle}$ exhibits a pronounced dependence on both the xc-functional and the spin multiplicity, as summarized in Table~\ref{tab:split-JT}.
Using semilocal functionals, namely the local-density approximation (LDA)~\cite{Perdew1981LDA} and PBE, we obtain the triplet $D_{2d}$ configuration as the ground state. \chg{In contrast, HSE06 stabilizes a singlet state and consequently predicts the triplet configuration to be metastable.
When switching from the stable to the metastable configuration, the symmetry changes to $D_{2}$ in PBE and LDA calculations, whereas in HSE06 the structure remains in the $D_{2d}$ symmetry.
In the former cases, this behavior can be interpreted as a Jahn--Teller distortion: the singlet state with $D_{2}$ symmetry is about 0.45~eV lower in energy than its $D_{2d}$ counterpart, in good agreement with the value reported by Goss \textit{et al.}~\cite{Goss2001PRB}.}
It is noteworthy that in HSE06 calculations the $D_{2d}$ geometry is 0.27~eV lower in energy than the $D_{2}$ configuration.

\begin{table}[t]
\caption{Relative energies (eV) of the $I_1^{\langle001\rangle}$ configuration for different symmetries and
spin states obtained with LDA, PBE, and HSE06. The lowest-energy configuration for each functional is taken as the zero reference. \chg{The value reported in parentheses is calculated within the von Barth procedure.} }
\label{tab:split-JT}
\begin{ruledtabular}
\begin{tabular}{lrlll}
Sym.     & Spin & LDA & PBE & HSE06 \\ 
\hline
$D_{2}$  & singlet & 0.02 (0.04) & 0.33 (0.67) & 0.27(0.55)   \\
$D_{2}$  & triplet & -- & -- & ---    \\
$D_{2d}$ & singlet & 0.49 (0.97) & 0.79 (1.59) & 0.00   \\
$D_{2d}$ & triplet & 0.00 & 0.00 & 0.02 (0.04) \\ 
\end{tabular}
\end{ruledtabular}
\end{table}

Our results also indicate that the bond-centered defect is 2.37 eV higher in energy than the more stable $I_1^{\langle001\rangle}$ configuration.
This defect with a $D_{3d}$ symmetry stabilizes in a triplet spin state while having a singlet state at 0.85~eV higher energy, in agreement with the previous reports~\cite{Bernholc1988PRL, Weigel1973, Breuer1995}.

Three 2C$_i$ configurations, known as $\pi$-bonded, Humble, and di-$\langle 001\rangle$ split-interstitial ($I_2^{\langle001\rangle}$), shown in Fig.~\ref{fig:geo_inter} as \di{a}-\di{c}, comprise our dataset, listed in order of increasing energy.
Further structural details are also provided in Refs.~\cite{Goss2001PRB, Goss2001DRM, Twitchen1996-R1_EPR}.
The $\pi$-bonded di-interstitial, formed by two bond-centered mono-interstitials on opposite sides of the same hexagonal ring, exhibits $C_{2h}$ symmetry and a singlet ground state.
The Humble configuration consists of two \(I_1^{\langle001\rangle}\) units occupying \emph{next-nearest-neighbor} sites.
This exhibits $C_{2v}$ symmetry and a singlet ground state; the triplet state lies 0.13~eV higher in energy.
Notably, the positively charged Humble defect has been proposed to be related to the 3H optical center, with a ZPL energy of 2.46~eV~\cite{Goss2001PRB}. \chg{However, our findings indicate that such a conclusion about the charge state is incorrect because that charge state is not photostable under blue emission (see Sec.~\ref{subsect:formation_energy}). We rather identify 3H optical center with the \emph{neutral} Humble (\di{b}) defect in Sec.~\ref{sect:3h_center}.}
Finally, the $I_2^{\langle001\rangle}$ consists of two \(I_1^{\langle001\rangle}\) defects occupying \emph{nearest-neighbor} sites, exhibits $C_{2h}$ symmetry, and has historically been attributed to the triplet-state EPR center R1~\cite{Twitchen1996-R1_EPR, Komarovskikh2024}.
In agreement with prior works, we recover the same stability ordering: the $\pi$-bonded configuration is lowest in energy; the Humble one is 0.84~eV higher; and the $I_2^{\langle001\rangle}$ is 1.40~eV higher, all relative to the $\pi$-bonded configuration.

We identify five different tri-interstitial complexes (Fig.~\ref{fig:geo_inter}\tri{a}–\tri{e}); to our knowledge, the structures \tri{a}–\tri{d} have not been reported previously.
Configuration \tri{a}, with symmetry of $C_{1}$,  consists of a triangular moiety with one bond to a Humble-like di-interstitial [cf. Fig.~\ref{fig:geo_inter}\di{b}], and the third vertex is capped by a C$-$C bond that links to the opposite side of the Humble unit.
Note that an analogous triangular tri-interstitial motif has been discussed in silicon, so-called Coomer~\cite{Coomer1999}.
Complex \tri{b} can be viewed as comprising four split-interstitial units (cf. \mono{a}), three of which are nearly aligned along the $\langle010\rangle$-orientation, while the fourth is tilted.
Alternatively, it can be described as two dihedral motifs with internal angles of 116.44$^{\circ}$ and 118.94$^{\circ}$, related by a rotation about the [010]-axis.
This configuration has a $C_{1}$ symmetry and lies 1.46~eV higher in energy than the most stable tri-interstitial structure.
Defect \tri{c} also adopts a Coomer-like geometry, forming a highly compact complex that induces substantial local distortion in the crystal lattice~\cite{Coomer1999, Jones2002}. 
This configuration exhibits $C_{3v}$ symmetry and is energetically close to \tri{b}, lying 1.47~eV above the most stable tri-interstitial structure.
Configuration \tri{d} consists of a Humble-like di-interstitial [cf. \di{b}] adjacent to a Coomer-like motif [cf. \tri{c}], bridged by a single additional carbon atom.
It has a $C_{1}$ symmetry.
Energetically, configuration \tri{d} lies 2.90~eV above the most stable tri-interstitial.
It is structurally analogous to \tri{a}, but the capping carbon dimer is replaced by a single carbon atom.
The final tri-interstitial configuration considered, \tri{e}, has $D_{2d}$ symmetry and lies 3.48~eV above the most stable tri-interstitial structure.
We note that in Ref.~\cite{Goss2001DRM} this configuration was reported as the lowest-energy tri-interstitial.
In our work, configuration \tri{e} is recovered: placing four split-interstitials on sites adjacent to a vacancy reconstructs into this complex.
All tri-interstitial configurations are found to be non-magnetic, a point that will be further discussed in a later section.
It is worth noting that the structure reported in Ref.~\cite{Goss2001PRB,Goss2001DRM} was also identified in our searches; however, it lies outside the considered energy range of this study, being more than 4 eV higher in energy than the most-stable configuration.

Our calculations reveal two distinct tetra-interstitial complexes.
Configuration \tetra{a}, the well-known platelet defect~\cite{Fallon1995}, has $D_{2d}$ symmetry and consists of four $I_1^{\langle010\rangle}$ split interstitials that fully reconstruct, eliminating all dangling bonds.
This structure has been proposed as the fundamental building block of one-dimensional platelets in type-I diamond, where nitrogen is present as an impurity~\cite{Goss2000JPCM,Goss2003}.
Platelets themselves are extended, planar defects oriented along a single crystallographic direction.
Configuration \tetra{b} consists of two parallel di-interstitial \di{c} units and lies 1.77~eV higher in energy than the platelet configuration \tetra{a}.
Both tetra-interstitial complexes are non-magnetic. 

For the penta-interstitial, we identify two novel configurations.
Configuration \penta{a} consists of five aggregated split-interstitial units, has crystal symmetry $C_{2}$, and a triplet ground state. \chg{The singlet spin-state is nearly degenerate with the triplet state, lying 5 meV lower in energy.} 
Configuration \penta{b} lies 3.91~eV higher in energy than \penta{a}; it is oriented perpendicular to the (111) plane and is formed by a zigzag stacking of five carbon interstitials occupying hexagonal and tetrahedral interstitial sites. This defect has a $C_{1h}$ symmetry.

Finally, we identify two novel hexa-interstitial configurations, both formed through the aggregation of split-interstitial units.
Configuration \hexa{a} induces a pronounced lattice distortion and exhibits $C_{1}$ symmetry.
This configuration is very similar to structure \penta{a}; however, the presence of an additional split-interstitial, arranged so as to break one of the symmetry planes, lowers the symmetry.
\chg{Here, the singlet and triplet spin states are nearly degenerate, with the singlet lying within 1 meV below the triplet.} 
Configuration \hexa{b}, which at first glance resembles three parallel di-interstitials of type \di{b}, in fact consists of a combination of one \di{b} di-interstitial and one \tetra{a} tetra-interstitial unit. Two dangling bonds are present in the \di{b}-like Humble segment of this complex. This configuration lies 3.18~eV higher in energy than the most stable hexa-interstitial structure. \chg{The results discussed above are summarized in Table~\ref{tab:singlet-triplet}. However, singlet--triplet splittings below 10~meV remain inconclusive for identifying the true ground state within the numerical and methodological uncertainty of our approach.}
\begin{table}[htb!]
\caption{The spin multiplicities of defects in the neutral charge state are listed according to the labels used in this work and their conventional names, found in the literature. The triplet-singlet energy difference ($\Delta E_{TS}$) is reported in meV, with negative value indicating a triplet ground state.}
\label{tab:singlet-triplet}
\begin{ruledtabular}
\begin{tabular}{llcc}
Defect   & labels  & spin ground state & $\Delta E_{ST}$ \\ 
\hline
 $\mathrm{C}^{a}_{\mathrm{1i}}$ & $\langle 001\rangle$-split & singlet     &  40  \\
 $\mathrm{C}^{b}_{\mathrm{1i}}$ & Bond-centered              & triplet     & $-850$    \\
 $\mathrm{C}^{a}_{\mathrm{2i}}$ & $pi$-bond                  & singlet     & ---   \\
 $\mathrm{C}^{b}_{\mathrm{2i}}$ & Humble                     & singlet     & 130   \\ 
 $\mathrm{C}^{c}_{\mathrm{2i}}$ & di-$\langle 001\rangle$-split& triplet   & $-60$    \\ 
 $\mathrm{C}^{a}_{\mathrm{5i}}$ & $\mathrm{C}^{a}_{\mathrm{5i}}$ & singlet & 5   \\
 $\mathrm{C}^{b}_{\mathrm{5i}}$ & $\mathrm{C}^{b}_{\mathrm{5i}}$ & singlet & 140   \\ 
 $\mathrm{C}^{a}_{\mathrm{6i}}$ & $\mathrm{C}^{a}_{\mathrm{6i}}$ & singlet & 1   \\
 $\mathrm{C}^{b}_{\mathrm{6i}}$ & $\mathrm{C}^{b}_{\mathrm{6i}}$ & singlet & 60   \\
\end{tabular}
\end{ruledtabular}
\end{table}

\subsection{Bond Analysis}
\label{subsect:bond_analysis}

\begin{figure*}[t!]
\centering
    \includegraphics[scale=.95]{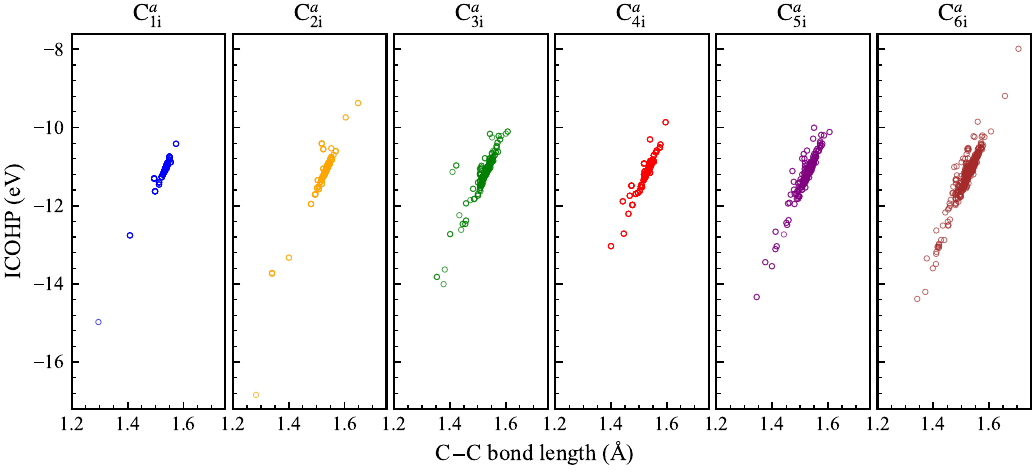}
    \vspace{-2mm}
    \caption{\label{fig:lobster_icohp} The integrated crystal orbital Hamilton population (ICOHP) values for the most stable of each interstitial defect versus their bond lengths. Each data point represents one C$-$C bond.}
\end{figure*}
Figure~\ref{fig:lobster_icohp} shows the dependence of the ICOHP on the C$-$C bond length for the lowest-energy mono- to hexa-interstitial configurations, \mono{a}-\hexa{a}.

For pristine diamond, the HSE06 calculations yield an ICOHP value of $-11.4$ eV, whereas PBE gives a value of $-9.64$ eV, consistent with the prior report~\cite{Gorne2019}. Notably, even in an ideal crystal, bonds with nearly identical geometric lengths can display a small spread in ICOHP, reflecting the fact that ICOHP is governed not only by interatomic distance but also by the local electronic environment and the underlying orbital interactions.
Unless stated otherwise, all defect configurations are analyzed at the HSE06 level.

For the \mono{a}, most C$-$C bonds lie within a narrow range of 1.50-1.55~\AA.
A distinct double-bond-like feature appears at a shorter bond length of 1.29~\AA~ with an ICOHP of $-14.97$~eV; this bond is associated with the C atoms forming the $I_1^{\langle001\rangle}$ configuration. The next isolated data points correspond to the second most negative ICOHP value of $-12.75$~eV and, in fact, represent four equivalent bonds that connect the defect center to the diamond lattice.

For configuration \di{a}, the C$-$C bond lengths are predominantly distributed in the range $1.49$-$1.56$~\AA.
In this structure, only the central $\pi$ bond, with a length of 1.28~\AA, and the four ancillary bonds linking the neighboring C atoms to the two $\pi$-bond-forming carbons, with lengths of 1.33~\AA, exhibit pronounced shortening. Together, these two groups of bonds, characterized by ICOHP values of $-16.8$~eV and $-13.7$~eV, define the defect cluster.
Embedding this cluster in the diamond lattice induces both compression and stretching of hosting bonds, giving rise to shorter bonds down to 1.40~\AA{} and elongated bonds exceeding 1.60~\AA.

In configuration \tri{a}, the C$-$C bond lengths are narrowly distributed within the $1.50$-$1.56$~\AA~ range, indicating strongly localized covalent bonding. Within the defect core, however, two significantly shortened bonds emerge at 1.35~\AA~and 1.37~\AA. Notably, the more negative ICOHP value corresponds not to the shortest bond, but to the second-shortest one (1.37~\AA).
The resulting contraction of the defect core forces the surrounding lattice to relax outward, producing neighboring C$-$C bonds that are elongated relative to the pristine lattice and appear in the upper (less negative ICOHP) region of the plot.

For the larger clusters, \tetra{a}-\hexa{a}, the dispersion in ICOHP broadens with increasing cluster size. This enhanced spread indicates the coexistence of comparatively strong and weak C$-$C interactions within a single defect complex, reflecting the increasing structural and bonding diversity as additional interstitial carbon atoms are incorporated.
Across all panels, shorter C$-$C bonds are generally associated with more negative ICOHP values, indicating stronger bonding, although a modest scatter is present and occasional deviations occur depending on the local coordination and electronic environment. Overall, the figure supports ICOHP as a reliable descriptor of C$-$C bond strength across the full range of interstitial cluster sizes.

Bader charge analysis~\cite{Henkelman2006} shows that, across all configurations from \mono{a} to \hexa{a}, a charge transfer of roughly 1.5~electrons occurs among the defect carbon atoms, with similar redistributions observed in the remaining systems. This systematic charge imbalance introduces a polar component into the core C$-$C bonds, partially shifting their character away from purely covalent. The resulting polarization enhances the local electrostatic attraction between the involved atoms, thereby strengthening these bonds relative to the surrounding, more covalent lattice bonds. Consequently, the bonds experiencing the largest charge transfer consistently correspond to the most negative ICOHP values within each defect structure, reflecting their increased bond strength.

\subsection{Formation Energies}
\label{subsect:formation_energy}

\begin{figure*}[ht!]
    \includegraphics[scale=.9]{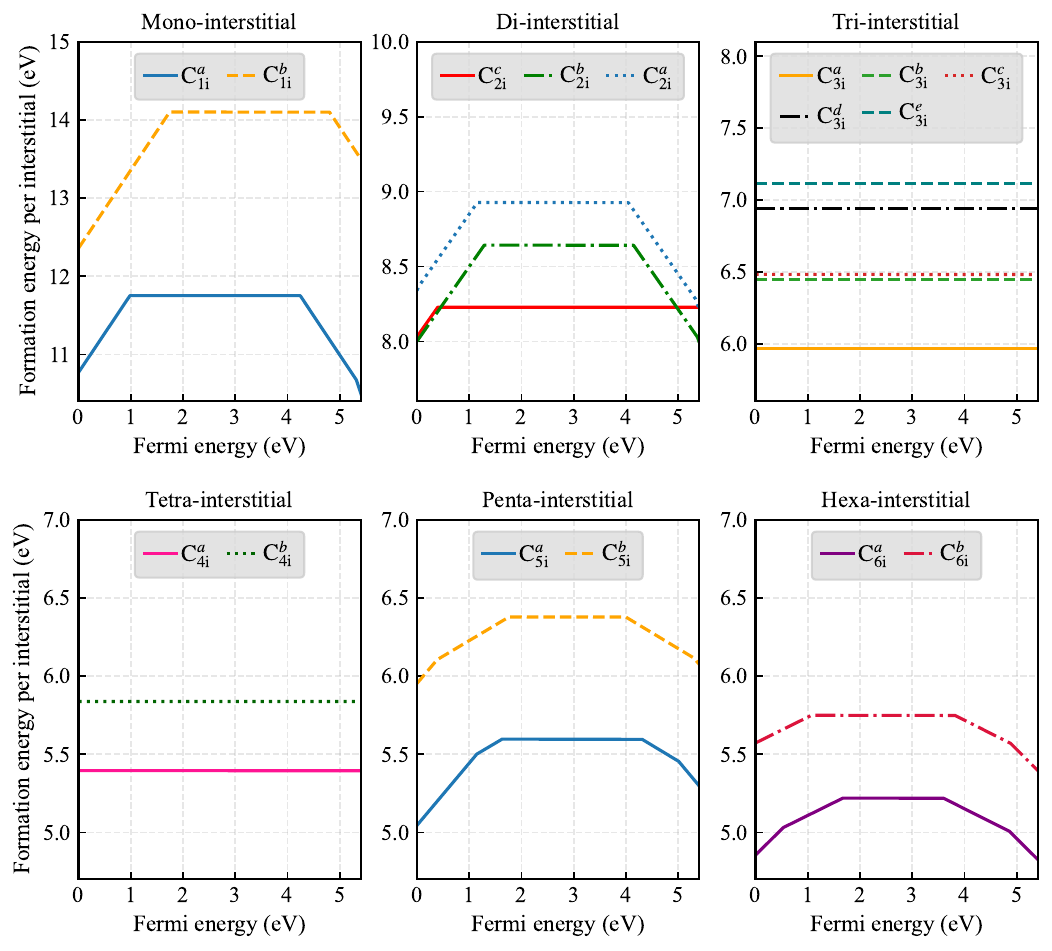}
    \centering
    \caption{\label{fig:formation} 
    Formation energy of the interstitial defects depicted in Fig.~\ref{fig:geo_inter} as a function of the position of the Fermi level. The valence level energy is aligned to zero for the sake of simplicity. The conduction level energy is set at 5.4 eV. }
\end{figure*}

We calculated the formation energy per interstitial of the optimized defects (cf. Fig.~\ref{fig:geo_inter}) to show their (relative) stability in different charge states (see Fig.~\ref{fig:formation}).

For the mono-interstitial, both configurations can exist in positive and negative charge states.
In the neutral charge state, configurations \mono{a} and \mono{b} have formation energies of 11.7 and 14.1 eV, respectively.
Although p-type or n-type doping shifts the Fermi level and thus can modify the preferred charge state, we find that the neutral defect remains thermodynamically stable over a wide range, indicating that it is not easily ionized under typical conditions.
Generally, defect charging lowers the formation energy when electron exchange with the Fermi reservoir allows the defect to relax into a lower-energy electronic and structural configuration.

For the di-interstitial complexes, configuration \di{a} is stable only in neutral and positive charge states, indicating that it has no empty defect level in the band gap that could accommodate an additional electron.
In the neutral state, configurations \di{a}, \di{b}, and \di{c} have formation energies of 8.2, 8.6, and 8.9 eV per interstitial, respectively.
By contrast, configurations \di{b} and \di{c} are stable in positive, neutral, and negative charge states.
At low and high Fermi-level energies, the positively and negatively charged states of the \di{b} defect, respectively, are the most stable among the di-interstitial configurations considered, consistent with previous findings~\cite{Goss2001PRB}.
\chg{We note that these defects are photostable under visible illumination only in the neutral charge state: both the positive and negative charge states are photo-converted to neutral via single-photon absorption for photon energies above 1.4~eV. Consequently, the assignment of the positively charged \di{b} defect to the 3H optical center proposed in Ref.~\cite{Goss2001PRB}, which is excited in the blue, is no longer supported.}

All tri-interstitial defects are found to be stable only in the neutral charge state.
The calculated formation energies per interstitial are 5.96, 6.44, 6.47, 6.94, and 7.12 eV for configurations \tri{a}, \tri{b}, \tri{c}, \tri{d}, and \tri{e}, respectively.
The absence of stable charged states indicates that these complexes do not introduce partially filled defect levels within the band gap, i.e., no dangling-bond-derived states are present.
Consequently, no charge-transition levels appear in the gap.
If all defect-related electronic states lie deep below the valence band maximum or high above the conduction band minimum, they remain inactive within the accessible Fermi level range.
This implies that the tri-interstitial defects cannot be ionized and thus remain electronically neutral irrespective of p-type or n-type doping.

The tetra-interstitial defects exhibit the same behavior as tri-interstitials: they are stable only in the neutral charge state.
The calculated formation energies per interstitial are 5.40~eV for configuration \tetra{a} and 5.84~eV for configuration \tetra{b}.

For the penta-interstitials, configuration \penta{a} is stable in charge states ranging from doubly positive to doubly negative.
The presence of dangling bonds in this complex gives rise to several occupied and empty defect levels within the band gap.
Electrons can be added to the empty levels or removed from the occupied ones, which leads to pronounced multi-charge state behavior in the formation energies as a function of the Fermi level.
A similar situation occurs for configuration \penta{b}, which is stable in the doubly positive, singly positive, neutral, and singly negative charge states.
In the neutral charge state, the formation energies per interstitial are 5.60~eV for \penta{a} and 6.38~eV for \penta{b}.

For the hexa-interstitial complexes, both configurations exhibit multiple stable charge states, similar to the penta-interstitials.
In the neutral state, configurations \hexa{a} and \hexa{b} have formation energies per interstitial of 5.22 and 5.75~eV, respectively.
Among all defects studied, \hexa{a} shows the narrowest Fermi level range over which the neutral charge state is thermodynamically stable, implying that even modest Fermi level shifts (e.g., due to doping) will drive it into a charged state.
Thus, \hexa{a} is the most easily ionized complex in our set.

We predict an overall trend in which the formation energy per interstitial atom decreases as the number of interstitials in the complex increases, with a pronounced drop from the single and di-interstitials to the tri- and tetra-interstitial complexes. 
This behavior indicates a thermodynamic driving force toward aggregation, consistent with earlier reports on the relative stability of extended self-interstitial complexes in diamond~\cite{Goss2000JPCM, Goss2003}.
Although this does not imply spontaneous clustering under all conditions, it shows that, once several interstitials are present, compact configurations such as the tetra-interstitial platelet \tetra{a} are energetically more favorable than an equivalent number of isolated split interstitials.

\subsection{Electronic Structure}
\label{subsect:electronic_props}

\begin{figure*}[htb!]
\centering
    \includegraphics[scale=.9]{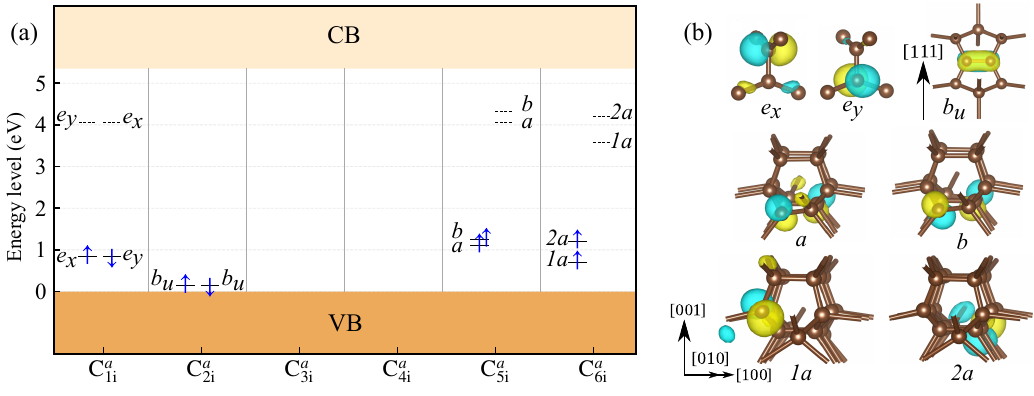}
    \caption{\label{fig:KS} (a) Electronic structure for the ground states of the neutral defects using HSE06 functional. Kohn-Sham (KS) energy levels are represented by spin-up ($\uparrow$) and spin-down ($\downarrow$). The valence band (VB) and conduction band (CB) are depicted in brown and cream, respectively. (b) Representation of the KS orbitals of the relevant states. The light cyan and yellow lobes exhibit negative and positive isovalues. The isosurface absolute value is set to $7 \times 10^{-7}$ {\AA}$^{-3}$. }
\end{figure*}
The ground state electronic structures of all the most stable interstitial defects (\mono{a}-\hexa{a}) are summarized in Fig.~\ref{fig:KS}(a).
The corresponding orbitals are visualized in Fig.~\ref{fig:KS}(b).

The mono-interstitial adopts a singlet ground state with a wave function of $D_{2d}$ symmetry.
Its $a_1$ state is fully occupied and lies within the valence band (VB).
In this case, the presence of two dangling bonds introduced by the $p$-orbital of the two $sp^2$ coordinated neighboring carbon atoms gives rise to a pair of degenerate $e$ levels in each spin channel within the band gap.

We illustrated the orbital localization of the defect, in which the occupied $e_x$ state is highly localized on the upper atom of the split interstitial with $p$-like character. We observe a small contribution from the neighboring atoms. This confirms that the $e_x$ state is a defect localized pure $p$ character, with nearly equal projection onto the $p_x$ and $p_z$ orbitals in the global frame and negligible $s$ or $d$ admixture. 
In addition, the unoccupied $e_y$ level shows an analogous localization pattern on the other atom of the defect. Here again, most of the weight resides on this atom, with almost equal $p_x$ and $p_z$ contributions and only small $s/p$ tails on the adjacent neighbors.

In an ideal symmetric singlet, one electron occupies each spin channel, distributed equally between the two components of the $e$ manifold.
This behavior is notably reproduced by the PBE functional: the two KS levels remain strictly degenerate, and each carries an occupation of $\approx 0.5$ per spin channel, corresponding to a half-filled $e$ doublet with the correct singlet occupation.
In contrast, the HSE06 hybrid functional lifts this degeneracy.
Although the two states still transform as an $E$ representation, one component ($e_x$ in spin-up and $e_y$ in spin-down) is lowered in energy and becomes fully occupied, while the complementary component ($e_y$ in spin-up and $e_x$ in spin-down) is shifted upward by roughly 3 eV and remains empty.
Thus, HSE06 drives the system from the symmetric half-half occupation characteristic of PBE to an integer-occupation solution in which one member of the $e$ pair is filled, and the other is depopulated.
\chg{We emphasize that} the KS level splitting is not a consequence of a Jahn–Teller distortion nor of any underlying lattice-symmetry breaking; rather, it reflects the hybrid functional's intrinsic tendency to penalize fractional occupations, together with the increased level of separation introduced by Fock-exchange.
\chg{This electron configuration instead corresponds to one of the Slater determinants of the $^1B_1$ multiplet~\cite{Goss2001PRB}.
Promoting a spin-down electron from $e_y$ to $e_x$ yields a mixture of the $^1A_1$ and $^1B_2$ multiplet states, which are optically inactive in the high-symmetry configuration but may become dynamically activated through coupling to symmetry-distorting phonons~\cite{Goss2001PRB}.
Possible correlations of this defect with near-infrared and ultraviolet optical centers are discussed in Ref.~\cite{Goss2001PRB}.
Since the main focus of this work is to understand the properties of carbon self-interstitial clusters, we continue our investigation with the di-interstitial complexes.}

In the di-interstitial, with a singlet ground state, the defect wave function exhibits $C_{2h}$ symmetry.
Deep defect levels appear in the valence band, including \textit{b$_g$} and \textit{b$_u$} states located about 2.5 eV below the VBM energy. Within the band gap, we find only a single fully occupied \textit{b$_u$} level, with no corresponding empty state available for electronic promotion.
The associated KS orbital shows that this in-gap \textit{b$_u$} state is strongly localized on the two interstitial atoms forming the short bond.
The $b_{u}$ state is a localized $p$–$p$ bonding orbital of $\pi$-symmetry between the two interstitial atoms.
In other words, the in-gap level corresponds to a defect-localized $\pi$ bond constructed predominantly from the $p$-orbitals of the di-interstitial.

For the tri-interstitial defects, neither the most stable configuration nor the metastable ones introduce electronic states in the band gap.
These configurations are therefore \emph{silent} (i.e., electrically inert) defects: they do not generate in-gap states and thus exhibit no electrical or optical activity.
However, the stable defect with $C_1$ symmetry possesses a deep, localized $a$ state in the valence band, while no corresponding resonant state is found in the conduction band.

For the tetra-interstitial defects, in both stable and metastable configurations, we do not observe any states in the band gap, as all bonds are saturated.
In the platelet defect, consisting of an array of I$_4$ units, all the carbon atoms are fourfold coordinated so no $p$-orbitals appear due to threefold coordinated sp$^2$ carbon atoms.
This defect exhibits two levels, $a_1$ and $b_2$, both of which lie deep within the valence band.
Consequently, this defect is optically and electronically inactive, i.e., a silent defect.
Goss \textit{et al.}~\cite{Goss2001PRB} previously reported only the $a_1$ state of this defect, but because they could not accurately determine the band edges by semilocal DFT functionals, they assumed that this level lies in the band gap.
In contrast, our results show that this state, together with an additional defect state, lies deep within the valence band.

The most stable penta-interstitial defect has a triplet ground state, and the wave functions of the localized states exhibit $C_2$ symmetry.
The spin-up channel contains fully occupied $a$ and $b$ states, whereas the corresponding states in the spin-down channel are empty.
The orbital localization for both the $a$ and $b$ states shows characteristic localization on the p-orbitals of the $sp^2$ coordinated C-atoms, similar to that of a split-interstitial carbon defect. However, they are spatially separated on opposite sides of the complex.

The hexa-interstitial configuration has a triplet ground state, with two fully occupied $a$ states in the spin-up channel and two empty $a$ states in the spin-down channel. 
The corresponding defect orbitals exhibit a localization pattern similar to that of the penta-interstitial configuration.

We note that, in the interstitial defects, gap states arise whenever there are $sp^2$-hybridized carbon atoms with unsaturated $p_z$ orbitals. In contrast, the tri- and tetra-interstitial configurations contain only $sp^3$-hybridized carbon atoms and therefore lack $p_z$ orbitals that could generate localized states in the band gap. 
If the surrounding $sp^3$ bonds were significantly weakened, their associated defect levels would be expected to fall into the gap. However, as shown in Sec.~\ref{subsect:bond_analysis}, the tri- and tetra-interstitials actually exhibit stronger $sp^3$ C$-$C bonds than pristine diamond. 
Consequently, the corresponding defect states remain resonant with the bulk bands rather than forming isolated gap levels. By contrast, in the mono-, di-, penta-, and hexa-interstitials, the presence of $sp^2$ carbon leads to $p_z$-derived defect states within the diamond band gap.

\chg{All the most stable clusters that introduce mid-gap states and undergo electron excitation to the conduction band lead to bound-exciton excitations. These transitions typically occur in the ultraviolet regime; therefore, such defects are not considered in the present study.}
Based on our KS level analysis, none of the lowest-energy carbon self-interstitial \chg{cluster} configurations exhibits the electronic structure required to account for the TR12 center. 
The TR12 center has an experimental ZPL of $\sim$2.63~eV, implying that any viable defect candidate must possess two well-isolated defect states whose KS energy separation exceeds this value.
No such pair of in-gap states is found for the neutral self-interstitial \chg{cluster} configurations considered here.
Although their formation energies suggest that these lowest-energy defects can exist in multiple charge states, which is interesting in its own right, we do not pursue them further here. 

\subsection{Vibrational Properties}
\label{sect:vibrational_properties}

\begin{figure*}[ht!]
\centering
    \includegraphics[scale=0.8]{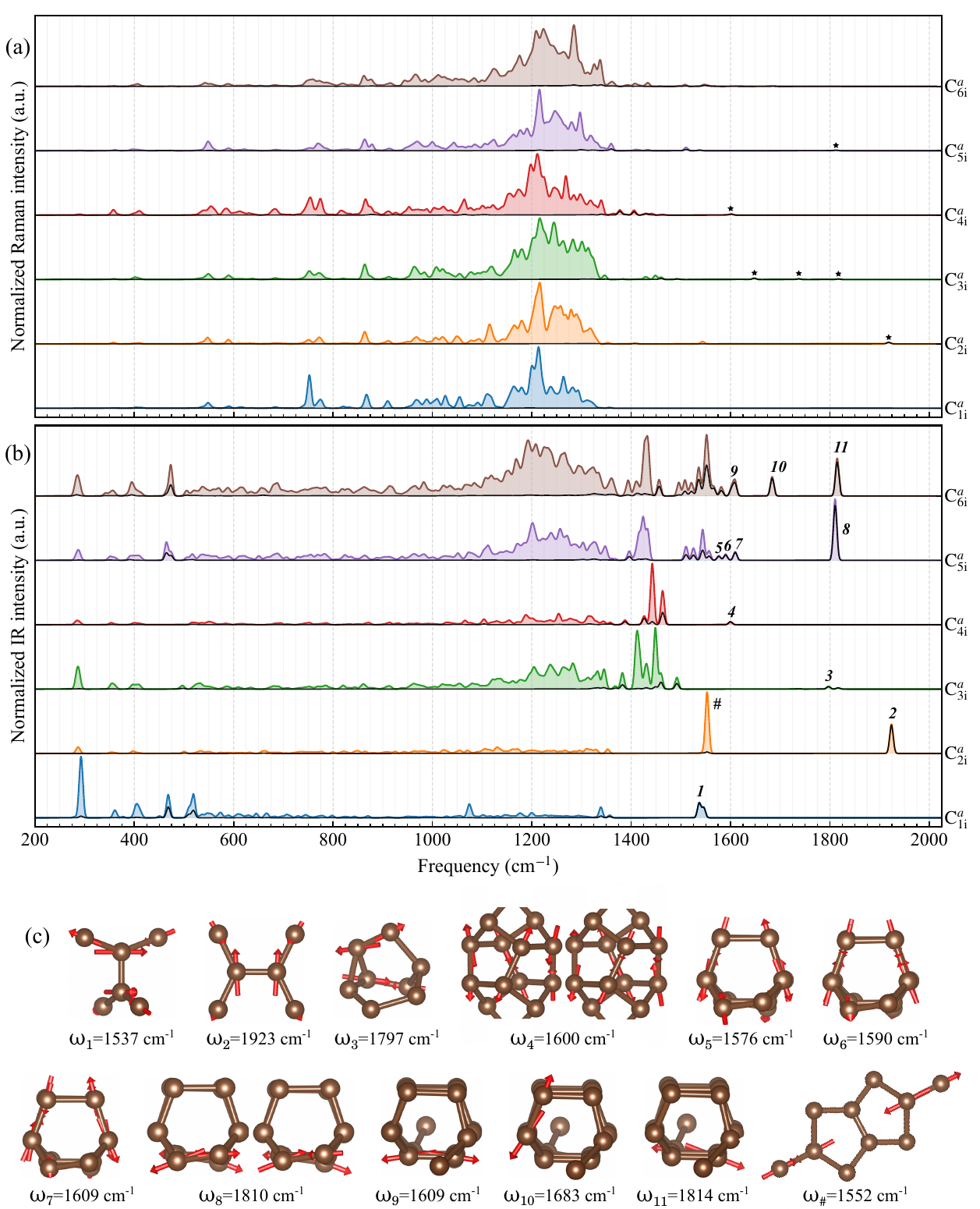}
    \caption{\label{fig:ir}
    (a) First-order Raman spectra for defect complexes \mono{a} to \hexa{a}.
    Pentagram symbols mark Raman-active modes that are predominantly localized on the defect centers.
    Each spectrum is normalized to its own maximum (arbitrary units). The black line shows the contribution of the defect atoms to the total spectrum, which are eligible in Raman cases.
    (b) Infrared (IR) spectra of the corresponding defects, where pronounced defect-localized modes are numbered and visualized in (c).
    (c) Atomic displacement patterns of the selected IR-active modes, where only the defect atoms are active. Red arrows indicate the direction of atomic motion, and their lengths are scaled by the displacement magnitude.}
\end{figure*}
For pristine diamond, the non-resonant first-order Raman spectrum shows a single, sharp peak at 1300 cm$^{-1}$, which is slightly lower than the experimental value of 1332 cm$^{-1}$, corresponding to the triply degenerate zone-center optical phonon characteristic of the $sp^{3}$-bonded lattice~\cite{Prawer1998_DRM}.
In contrast, no features appear in the infrared spectrum because all zone-center optical modes are IR inactive (silent) in diamond.
Note that the small deviation between the experiment and theory can be attributed to the choice of functional used in our vibrational calculations~\cite{El-Azhary_jpc_1996}.

Self-interstitials introduce local lattice disorder and thereby shorten phonon lifetimes.
In vibrational spectroscopy, this is presented as broadening, frequency shifts, and the appearance of additional peaks in the line shapes~\cite{Srivastava_Phonons_2022}.
Overall, the breakdown of translational symmetry relaxes the momentum selection rule, enabling phonons with finite wave vectors to contribute to features that are nominally associated with $\Gamma$-point active modes.
These effects give rise to defect-induced signatures in both Raman and IR spectra, which provide a sensitive fingerprint of the presence and nature of lattice imperfections.

The interstitial defects exhibit two distinct structural behaviors: in some configurations (e.g., structures \mono{a}, \di{a}, \penta{a}, and \hexa{a} in Fig.~\ref{fig:geo_inter}), the introduction of additional carbon atoms gives rise to dangling bonds, whereas in others (e.g., structures \tri{a} and \tetra{a}) the surrounding lattice reconstructs such that no explicit dangling bonds are present.
In both classes, our results show that the defects can nonetheless induce the formation of local $sp^{2}$-like bonding within the diamond lattice.
It is noteworthy that a change in the chemistry of the bonds can modify the IR activity of a mode, as the latter depends on changes in the dipole moment.
In particular, $sp^{2}$-hybridized carbon has a higher effective electronegativity than $sp^{3}$-hybridized carbon owing to its larger $s$-character, which alters the local charge distribution and, consequently, the associated dipole moment.

As shown in Fig.~\ref{fig:lobster_icohp}, self-interstitials give rise to bond lengths of 1.28–1.40~\AA, up to $\sim$17\% shorter than in pristine diamond (1.54~\AA), indicating locally stronger bonds and correspondingly higher-frequency vibrational modes. These bond lengths are comparable to those in molecular hydrocarbons, spanning a single C$-$C bond in ethane (1.54~\AA), a double C$=$C bond in ethene (1.34~\AA), and a triple C$\equiv$C bond in ethyne (1.20~\AA).
More specifically, the C$=$C and C$\equiv$C stretching modes occur at 1623 and 1974~cm$^{-1}$, respectively~\cite{shimanouchi1973tables}. This indicates that the additional high-energy features arise directly from interstitial-induced bonding, rather than phonon-folding.

Alongside the Raman and IR spectra shown in Fig.~\ref{fig:ir}, we also evaluated the contribution of the defect centers to the activity of each vibrational mode.
The contribution of the defect-center atoms to a given vibrational mode is quantified as
\begin{equation}
    \rho = \sum_{n} \sqrt{\left| v_{n,x} \right|^{2} 
                           + \left| v_{n,y} \right|^{2} 
                           + \left| v_{n,z} \right|^{2}},
\end{equation}
where $v_{n,\alpha}$ denotes the eigenvector component of the defect-center atom $n$ along the direction $\alpha$, which is a real number at the $\Gamma$-point.
Partial intensities are then obtained by scaling the total mode activities by the corresponding $\rho$ values.
Defect-center atoms are defined as those that are strongly displaced from their ideal crystallographic sites; their numbers are 6, 6, 8, 12, 14, and 18 atoms for the \mono{a} – \hexa{a} complexes, respectively.

Our results show that the Raman spectra in the lattice phonon region are broadly similar across all structures, whereas interstitial defects introduce several additional peaks.
A systematic blue shift is also observed, which we attribute to internal strain in the lattice.
In our calculations, the cell vectors are always fixed to those of the pristine crystal.
At the highest defect density, an average strain of about 1.5\% is expected; depending on the defect orientation, this strain can be uniaxial, biaxial, or triaxial.
From our calculations on compressively strained lattices, uniaxial, biaxial, and triaxial strains of 1.5\% shift the vibrational frequencies by approximately 31, 48, and 66~cm$^{-1}$, respectively.

Defects introduce both stiffer and softer bonds into the system.
Softer bonds could, in principle, contribute to Raman activity at low frequencies; however, our results show no localized defect-related contribution in this region.
In contrast, several high-frequency peaks appear between 1375 and 1925~cm$^{-1}$, which can be attributed to vibrations of the defect centers or to modes involving stiffer bonds between the defect centers and the host lattice.
Nevertheless, the corresponding non-resonant first-order Raman activities are negligible.

The IR spectra clearly exhibit many distinct peaks, indicating that IR spectroscopy is highly sensitive to self-interstitials and can reliably discriminate between different defect types.
Whereas a pristine diamond is IR-inactive at the zone center because inversion symmetry forbids dipole-active phonons, interstitial defects, especially those introducing local $sp^{2}$ bonding, break this symmetry and activate vibrational modes with nonzero dipole moments.
This symmetry breaking enables a range of IR-active features. Many of the stiffer, defect-induced bonds give rise to high-frequency IR modes with well-separated and clearly distinguishable peaks.
There are only four peaks at low frequency, localized on the defect; two at 470 and 518 cm$^{-1}$ in \mono{a}, and one in both \penta{a} and \hexa{a} at 465 and 474 cm$^{-1}$, respectively.
All high-energy IR-active modes are fully localized on the bonds associated with the defect centers, as illustrated in Fig.~\ref{fig:ir}(c).

\chg{
\subsection{Identification of 3H center and its relation with TR12 center}
\label{sect:3h_center}

\begin{figure}[t!]
\centering
\includegraphics[scale=.99]{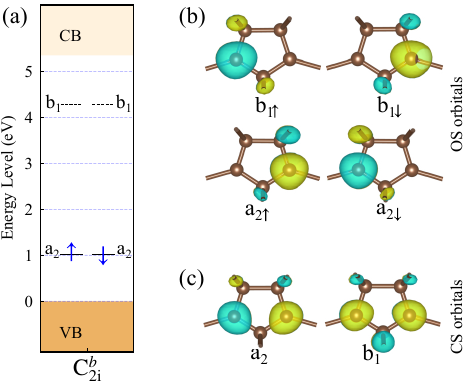}
\caption{(a) The Kohn–Sham levels of the open-shell singlet state of the neutral \di{b} defect, represented by spin-up ($\uparrow$) and spin-down ($\downarrow$) channels.
Representation of the defect-state orbitals for (b) an open-shell (OS) singlet broken-symmetry wave function and (c) a closed-shell (CS) symmetric wave function treatment. The light cyan and yellow lobes exhibit negative and positive isovalue, with an absolute value of $5 \times 10^{-7}$ {\AA}$^{-3}$.}
\label{fig:ks-humble}
\end{figure}
%
The experimental detection of the R1 EPR center, associated with the di-interstitial \di{c}, indicates that metastable configurations can likewise appear as local minima on the potential energy surface~\cite{Twitchen1996-R1_EPR, Twitchen1999_R1}.
To this end, we also consider configurations that can form and persist under kinetic control.
Among these, the \di{b} defect (the Humble di-interstitial) is particularly compelling because it introduces both filled and empty levels in the fundamental band gap of diamond.

As mentioned, Goss \textit{et al.}\ proposed that the 3H optical center corresponds to \di{b} in the positive charge state based on the computed local vibration modes that were compared to experiments~\cite{Goss2001PRB, Steeds1999}. Our formation-energy diagram showed that under optical excitation, the positively charged defect would ionize to the neutral charge state, so it would not remain stable at the relevant excitation energy.
Therefore, this attribution appears to be incorrect. On the other hand, the neutral charge state cannot be disregarded as a viable candidate which is photostable under blue illumination.

The neutral \di{b} defect forms a two-state system, as shown in Fig.~\ref{fig:ks-humble}(a). It exhibits a broken-symmetry singlet ground state, as seen from orbitals visualization in Fig.~\ref{fig:ks-humble}(b). In order to test this assertion, we performed symmetric closed-shell (CS) calculations (Fig.~\ref{fig:ks-humble}(c)), which show that its energy is 0.56~eV higher than that obtained from our OS singlet calculations.
In a broken-symmetry system, the wave functions spontaneously break the space-group symmetry, and the KS orbitals are not necessarily eigenstates of the symmetry operations of the defective supercell~\cite{Grafenstein2002}.
Such a dilemma can be alleviated by adopting a multiplet-state representation~~\cite{Thiering2015, Aleksei2023}. 
Considering the $a_{2}$ and $b_{1}$ defect orbitals in $C_{2v}$ symmetry, four states are obtained:
\begin{subequations}
\begin{align}
\Psi\!\left(^{1}A_1\right) &= 
\big[a_2(1)a_2(2)\big]
\big[\uparrow\downarrow-\downarrow\uparrow\big], \label{eq:oss}\\[2mm]
\Psi\!\left(^{1}A_1\right) &= 
\big[b_1(1)b_1(2)\big]
\big[\uparrow\downarrow-\downarrow\uparrow\big], \label{eq:ossex}\\[2mm]
\Psi\!\left(^{1}B_2\right) &= 
\big[a_2(1)b_1(2)+b_1(1)a_2(2)\big]
\big[\uparrow\downarrow-\downarrow\uparrow\big], \label{eq:ossgs}\\[2mm]
\Psi\!\left(^{3}B_2\right) &= 
\big[a_2(1)b_1(2)-b_1(1)a_2(2)\big]
\left\{
\begin{array}{c}
\big[\uparrow\uparrow\big] \\
\big[\uparrow\downarrow+\downarrow\uparrow\big] \\
\big[\downarrow\downarrow\big] \label{eq:ossexs}
\end{array}
\right\}.
\end{align}
\label{eq:wfs}
\end{subequations}

Equation~\ref{eq:oss} represents the CS singlet state. 
Equation~\ref{eq:ossex} corresponds to a double-electron excitation configuration, which lies $2.58$~eV higher in energy than the ground state. 
However, the relevant OS singlet ground state arises when one electron occupies the spin-up channel of the $a_{2}$ orbital and the other occupies the spin-down channel of the $b_{1}$ orbital (or vice versa), corresponding to the $a_{2}\otimes b_{1} = {}^{1}B_{2}$ configuration, as given in Eq.~\ref{eq:ossgs}. 
This state is described by a broken-symmetry wave function that reproduces the ground-state density of the multiplet state.
The corresponding triplet state, $^{3}B_{2}$, given by Eq.~\ref{eq:ossexs}, lies 0.13~eV higher in energy.

The relevant excitation of the 3H center transforms the ground state [Eq.~(\ref{eq:ossgs})] into the excited state [Eq.~(\ref{eq:ossex})] via a double-excitation process, yielding a ZPL of 2.58~eV, in good agreement with the experimental value of 2.46~eV~\cite{Steeds1999}.
A single-electron excitation cannot describe this transition, as it does not recover the multideterminant, spin-adapted singlet configuration required for an accurate many-body description. 
In principle, the two CS $^1A_1$ states should be treated as linear combinations (i.e., mixed). However, because the calculated energy separation between the configurations in Eqs.~(\ref{eq:ossgs}) and (\ref{eq:ossex}) is $\approx 2.0$~eV, we expect the mixing to be weak and therefore regard our approach as a reasonable approximation.

To compute the vibrational properties, we adopt the optimized excited-state geometry [Eq.~(\ref{eq:ossex})], which is well described as a single Slater-determinant CS configuration and can therefore be treated within the $\Delta$SCF framework using both hybrid and semilocal density functionals. This treatment is consistent with the Huang--Rhys approximation employed here to generate the optical spectra of the defects.
The calculated PL spectrum is shown in Fig.~\ref{fig:pl}. After applying a rigid shift of 0.12~eV to align the calculated spectrum with experiment, we reproduce the characteristic vibrational features of the 3H center. We obtain a total Huang--Rhys factor of $S=0.36$, corresponding to a Debye--Waller factor of 0.69 (69\%).
\begin{figure}[htb!]
\centering
\includegraphics[scale=1]{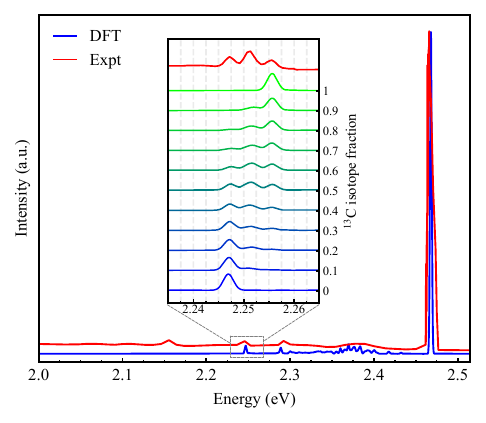}
\caption{The PL spectrum of the neutral \di{b} defect compared with the experimental data from Ref.~\cite{Steeds1999}. A rigid shift of $0.12$~eV is applied to the calculated spectrum to align the zero-phonon line with experiment. The inset highlights the evolution of the PL spectrum as a function of isotope mass. The experiments were carried out for a diamond sample with a 50\%-50\% mixture of $^{13}$C and $^{12}$C isotopes (red spectrum).}
\label{fig:pl}
\end{figure}

The PL spectrum of the 3H optical center was also measured in a specially prepared diamond sample containing a 50\%--50\% mixture of $^{13}$C and $^{12}$C isotopes~\cite{Steeds1999}.
This isotopic disorder gives rise to additional features in the PL spectrum; most notably, a pronounced triple peak emerges in the phonon sideband (PSB) around 2.25~eV, which has been associated with local vibrational modes of $\sim$230~meV.
Isotope-induced shifts and splittings in the PSB provide a sensitive means to validate microscopic models of optical centers. In addition to the energies of the local modes, their relative intensities in the PL spectrum are also informative, and are accessible within the Huang--Rhys framework used here.

Isotope enrichment systematically renormalizes vibrational frequencies: increasing the average atomic mass lowers phonon frequencies and compresses the PSB toward the ZPL.
To reproduce the experimentally observed triple-peak structure in the 2.24--2.26~eV range, we generated 50 random isotope configurations and report the ensemble-averaged lineshape.
Throughout, we assume that isotope substitution leaves the interatomic force constants essentially unchanged; accordingly, only the atomic masses were modified when constructing the dynamical matrix (see, e.g., Ref.~\cite{GhafariPRB_2025}).
The dynamical matrix was then solved separately for each configuration.

Mass substitution results in an approximately 10~meV shift of the PSB.
The characteristic triple-peak structure is most pronounced at a 50\% $^{13}$C concentration, yielding excellent agreement with experiment.
This agreement provides additional support for assigning the 3H center to the neutral \di{b} defect.
Finally, the experimentally observed peak at 2.15~eV, which is absent from our calculations, likely originates from residual laser light or a different color center rather than a 3H-related emission feature.

We note that the 3H center is particularly interesting because it features a singlet ground state and an experimentally confirmed photoactive singlet excited state~\cite{Steeds1999}, together with a low-lying triplet manifold. This level structure could be exploited for qubit operation, where the metastable triplet state may serve as an optically addressable qubit. However, the large energy separation between the optically excited singlet and the metastable triplet is expected to suppress intersystem crossing (ISC), potentially limiting spin polarization and readout. One possible route to enhance ISC is a two-color excitation scheme, in which a second laser resonantly drives vibronic transitions to access spin–orbit-coupled channels more efficiently, thereby enabling qubit operation.

\begin{figure*}[htb!]
\centering
\includegraphics[scale=1]{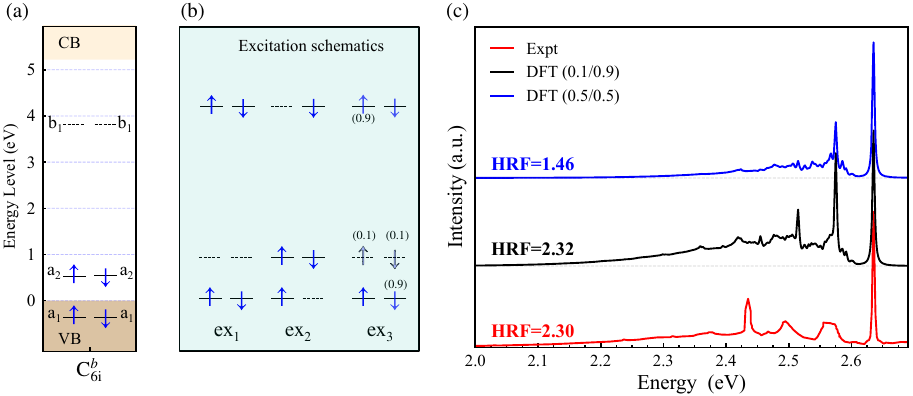}
\caption{
(a) The Kohn–Sham energy levels of the open-shell singlet state of the neutral \hexa{b} defect, shown separately for the spin-up ($\uparrow$) and spin-down ($\downarrow$) channels.
(b) Schematic illustration of the electron excitation mechanisms investigated in the search for the TR12 PL signal.
(c) The PL spectra of the neutral \hexa{b} defect compared with the experimental data taken from Ref.~\cite{Naydenov2009}.
For better comparison with the experiment, rigid energy shifts of $0.10$~eV and $-0.42$~eV were applied to the calculated spectra to align the ZPL with the experimental value. Our calculated ZPL for the occupation numbers of 0.1 and 0.5 are 2.53~eV and 3.06~eV, respectively.}
\label{fig:pltr12}
\end{figure*}
By establishing this identification and directly overlaying the experimentally measured PL spectra of the 3H and TR12 centers, several notable similarities become apparent.
Beyond the proximity of their ZPLs---2.46~eV for 3H and 2.63~eV for TR12---their phonon sidebands (PSBs) exhibit multiple common spectral features~\cite{Naydenov2009, Steeds1999}.
Furthermore, annealing studies show opposing trends in their intensities: the TR12 signal generally increases with annealing, whereas the 3H signal decreases~\cite{Iakoubovskii2000, Vlasov2002}.
This anticorrelation suggests that TR12 may form as a secondary product associated with the transformation of the 3H center.
Motivated by this possibility, we systematically screened our defect dataset for more complex configurations that intrinsically incorporate the Humble structural motif (i.e., \di{b}).
This search ultimately highlighted the \hexa{b} defect as a prime candidate.

The \hexa{b} defect is composed of the \di{b} and \tetra{a}. 
While the isolated \tetra{a} defect does not introduce localized states within the band gap and instead produces a deep valence-band state, it plays a crucial role when incorporated into the \hexa{b} configuration. Here we label the character of the electronic states as those for the isolated Humble \di{b} defect with $C_{2v}$ symmetry for the sake of simplicity; the symmetry of the composed defect is rather $C_{1h}$. 
As shown in Fig.~\ref{fig:pltr12}(a), the \tetra{a}-related state, labeled $a_1$, shifts toward the valence-band maximum and thereby may participate in the electronic excitation process.
When considering the $a_2$ and $b_1$ states, this configuration exhibits an electronic structure similar to that of the \di{b} defect, where the singlet ground state is described by a broken-symmetry solution. The singlet–triplet energy difference is 60~meV (see Table~\ref{tab:singlet-triplet}).

For two electrons occupying a four-state system, the many-body states can be constructed in the same manner as for the 3H configuration discussed above.
The broken-symmetry singlet ground state, $a_2\otimes b_1 = B_2$, is followed by the closed-shell $a_2\otimes a_2 = A_1$ state, which lies 0.73 eV higher in energy. The $^{3}B_2$ triplet state is found at 0.06~eV above the broken-symmetry singlet state.
The $b_1 \otimes b_1 = A_1$ configuration arises from a double-electron excitation and lies 2.46~eV above the ground state.
However, incorporating the two extra states originating from \tetra{a} leads to a six-state system with 15 many-body configurations (six singlets and three triplets with nine components), reflecting more complexity.

To identify the relevant excitation process, we use the experimental ZPL (2.63~eV) and HRF (2.30) of the TR12 center as reference criteria.
As illustrated schematically (Fig.~\ref{fig:pltr12}), one of the lowest optical transitions, denoted ex$_1$, corresponds to a double-electron excitation from the $a_2$ to the $b_1$ level, yielding a computed ZPL of 2.46~eV and an HRF of 3.22.
Another excitation, ex$_2$, involves a single-electron promotion from the $a_1$ to the $b_1$ state, for which we obtain a ZPL of 3.52~eV and an HRF of 0.84.
Comparison of these two candidate excitations suggests that the system behaves as a strongly interacting two-electron two-hole manifold, in which the actual optical transition may arise from a mixture of the ex$_1$ and ex$_2$ configurations rather than from a pure single excitation.
To probe this possibility, we introduced fractional occupations of the two excited configurations as a proxy for a mixed many-body state (see ex$_3$ in Fig.~\ref{fig:pltr12}(b)).
Specifically, we considered fractional occupations of $(0.1/0.9)$ and $(0.5/0.5)$ for the two excitation components.

Figure~\ref{fig:pltr12}(c) shows the PL lineshapes.
For the $(0.5/0.5)$ occupation, we obtain a ZPL of 3.06~eV and an HRF of 1.46.
The $(0.1/0.9)$ occupation yields a ZPL of 2.53~eV with an HRF of 2.32, which is closer to the experimental values.
However, the corresponding PSB structure does not fully reproduce the experimental spectral features.
We attribute this discrepancy either to limitations of the computational methodology employed in this work or to finite-size effects arising from the relatively small supercell used for the six-interstitial defect.

We propose that \hexa{b} may represent a potential candidate for the TR12 center; however, a definitive assignment requires higher-level excited-state methods capable of simultaneously capturing electron–electron correlation and vibronic coupling in the excited state. Such an investigation using multireference many-body approaches is beyond the scope of the present work.
Nevertheless, based on the energetic ordering, ZPL alignment, and qualitative agreement of the PL spectral features, we conclude that the \hexa{b} defect remains a viable candidate for the TR12 center.
}

\section{Conclusions}
\label{sec:conclusion}

We have carried out a systematic investigation of carbon self-interstitial complexes in diamond, spanning from mono- to hexa-interstitial defects.
To overcome the severe configurational complexity that emerges for larger clusters, we employed interatomic potential-assisted structure searches, followed by full relaxation and energetics within DFT.
Accurate hybrid functional calculations were used to ensure reliable defect stability ordering and defect-level alignments.
This multiscale protocol enabled us to produce all established low-energy configurations for small interstitials and to identify several previously unreported tri-, penta-, and hexa-interstitial geometries.

Our results confirm that the neutral $\langle 001\rangle$-oriented split interstitial is lowest-energy mono-interstitial configuration, while the commonly reported lower-symmetry alternative arises from the limitations of semi-local functionals. 
For di-interstitials, we recover the known stability hierarchy, with the $\pi$-bonded complex as the ground state.
For higher-order complexes, we find a clear thermodynamic drive toward aggregation: the formation energy per interstitial decreases with cluster size, showing a pronounced drop from isolated and di-interstitials to compact tri- and tetra-interstitials. The tetra-interstitial platelet is predicted to be a particularly stable motif, consistent with its proposed role as a building block for extended $\{100\}$ planar platelets. As additional carbon atoms are incorporated, the bond-length window remains narrow, yet the spread in bond strengths increases, reflecting the coexistence of very stiff defect-core bonds and weaker defect–lattice linkages.

The electronic structures reveal two distinct classes of self-interstitial complexes.
Mono-, di-, penta-, and hexa-interstitials can host localized in-gap states and multiple charge states, associated with dangling bonds or partly reconstructed motifs.
In contrast, all low-energy tri- and tetra-interstitials are electronically silent; they introduce no gap states and remain stable only in the neutral charge state, \chg{highlighting their relevance as optically and electrically silent defects in irradiated and annealed diamond}.
Crucially, none of the \emph{lowest-energy} neutral self-interstitial complexes exhibits the pair of well-isolated in-gap states required to account for the TR12 center with its $\sim$2.63 eV ZPL.
\chg{This observation rules out the most stable self-interstitial aggregates as microscopic models of TR12.
Motivated by this discrepancy, we extended our analysis to metastable configurations that may form and persist under kinetic control.
Within this broader configurational landscape, our investigation of metastable configurations identifies the 3H defect center as a neutral di-interstitial defect.
Building on this identification, we further propose that the TR12 center may originate from a 3H-containing defect linked to a metastable hexa-interstitial configuration.}

Vibrational spectroscopy provides complementary fingerprints of defects \chg{and might be the only experimental method for exploring electronically inactive defects}.
\chg{For the most stable complexes we analyze the Raman and infrared signatures associated with these defects.}
In Raman spectra, the lattice-phonon region remains broadly similar across defects, aside from systematic blue shifts originating from internal strain induced by fixed pristine lattice vectors.
Defect-center vibrations introduce additional Raman peaks at high frequency, but their non-resonant first-order activities are negligible.
By contrast, IR spectra show rich, defect-specific signatures. Local symmetry breaking, often accompanied by $sp^{2}$-like rehybridization, activates dipole-allowed modes that are absent in pristine diamond. Short, stiff C$-$C bonds within the defect cores yield well-separated high-frequency IR peaks (1375-1925 cm$^{-1}$), and mode-projection analysis confirms that these features are strongly localized on defect-center bonds. These results establish IR spectroscopy as a particularly powerful probe for discriminating among self-interstitial complexes.

Overall, this work delivers a unified and functionally robust picture of the stability, electronic activity, and vibrational signatures of carbon self-interstitial aggregates in diamond\chg{, and establishes a consistent framework for linking first-principles defect models with experimental identification.}

\begin{acknowledgments}
A. G.\ acknowledges support from the Quantum Information National Laboratory of Hungary, funded by the National Research, Development, and Innovation Office (NKFIH) under Grant No.\ 2022-2.1.1-NL-2022-00004, as well as the EU QuantERA II MAESTRO project (NKFIH Grant No.\ 2019-2.1.7-ERA-NET-2022-00045). A.\ G.\ further acknowledges access to high-performance computational resources provided by KIFÜ (Governmental Agency for IT Development, Hungary), and funding from the European Commission through the QuSPARC (Grant No.\ 101186889) and SPINUS (Grant No.\ 101135699) projects.

\end{acknowledgments}

\section{Data Availability}
The data that support the findings of this study are available in the ARP dataset
\href{https://doi.org/21.15109/ARP/5KBQCD}{ARP/5KBQCD}.

\bibliography{nima}
\end{document}